\documentclass[preprint,aps,prd,showpacs,nofootinbib,amsmath,amssymb,amsfonts,superscriptaddress]{revtex4-1}

\usepackage{amssymb}
\usepackage{bm}
\usepackage{latexsym}
\usepackage[normalem]{ulem}
\usepackage{soul}
\usepackage{amsfonts,amssymb}
\usepackage{graphicx,epsfig}
\usepackage{psfrag}
\usepackage{amsmath,amssymb}
\usepackage{mathrsfs}
\usepackage{amsthm}
\usepackage{bm}
\usepackage{lipsum}
\usepackage{array}
\usepackage{tabu}
\usepackage{latexsym}
\usepackage{mathrsfs}
\usepackage{hyperref}
\usepackage{float}
\usepackage[section]{placeins}
\hypersetup{
    colorlinks=true,       
    linkcolor=red,          
    citecolor=blue,        
}
\newcommand{\addlabel}[1]{%
Eq.\refstepcounter{equation}%
\addlabel{#1}%
\let\]\endequation }

\begin{document}

\title{Exact inflationary solutions in exponential gravity}

\author{Joseph P Johnson} \email{josephpj@iitb.ac.in}
 \affiliation{Department of Physics, Indian Institute of Technology Bombay, Mumbai 400076, India}
\author{Jose Mathew}\email{josecherukara@gmail.com} 
\affiliation{Department of Physics, Indian Institute of
Technology Madras, Chennai 600036, India}
  \author{S. Shankaranarayanan} \email{shanki@phy.iitb.ac.in}
 \affiliation{Department of Physics, Indian Institute of Technology Bombay, Mumbai 400076, India}

\begin{abstract}
We consider a modified gravity model of the form $ f(R,\phi)=R e^{h(\phi)R} $, where the strong gravity corrections are taken to all orders and $\phi$ is a self-interacting massless scalar field. We show that the conformal transformation of this model to Einstein frame leads to non-canonical kinetic term and negates the advantage of the Einstein frame. We obtain exact solutions for the background in the Jordan frame without performing  conformal transformations and show that the model leads to inflation with exit.  We obtain scalar and tensor power-spectrum in Jordan frame and show that the model leads to red-tilt. We discuss the implications of the same in the light of cosmological observations. 
\end{abstract}

\maketitle
\section{INTRODUCTION}

Cosmological inflation remains a successful paradigm during the high energy phase 
of the Universe~\cite{book:15483,book:15485,book:15486,book:75690,mukhanov1992theory}. The problems of the standard cosmology, like flatness/horizon, are elegantly solved if the comoving scales grow quasi-exponentially. The prediction of the simplest models of inflation is a flat Universe, i.e.
$\Omega_{\rm total} = 1$.  Importantly, it provides a natural mechanism for the generation of primordial density perturbations for understanding the initial conditions for structure formation and Cosmic Microwave Background (CMB) anisotropy. The metric perturbations at the last scattering surface are observable as the temperature anisotropy in the CMB~\cite{book:15483,book:15485,book:15486,book:75690,mukhanov1992theory}. 

Despite the simplicity of the inflationary paradigm, obtaining a canonical scalar field ($\phi$) driven inflationary model within General relativity(GR) requires highly fine-tuned potential~~\cite{book:15483,book:15485,book:15486,book:75690,mukhanov1992theory}.  Since inflation takes place at high energies, we need to consider quantum corrections to both gravity and matter sectors which result in modifications to general relativity~\cite{Donoghue:1993eb,Donoghue:1997hx,Hamber:1995cq,Barth:1983hb}.  A possible way to introduce modifications to GR is by adding higher order curvature terms like contracted Ricci/Riemann tensors or higher powers of Ricci scalar to the Einstein-Hilbert action~\cite{2015-Myrzakulov.etal-EPJC,Calcagni:2010ab}.  $f(R)$ models, where Lagrangian density is a general function of Ricci scalar, are the simplest among them \cite{sotiriou2010f,tsujikawa,clifton2011modified,nojiri2011unified}. These models are general enough to include the modifications to the GR in high energy limit, and more importantly, they do not suffer from Ostr\"ogradsky instability \cite{2007-Woodard-Proc}.

It has been shown that $f(R)$ models can successfully describe the inflationary phase of the universe \cite{tsujikawa}. However, the field equations associated with these models are higher order and non-trivial. To circumvent this problem, one performs conformal transformation so that the resultant action corresponds to Einstein-Hilbert action with a canonical scalar field. Although the two  --- original (Jordan frame) and the conformally transformed (Einstein frame) --- actions are related by conformal transformations, there are still some ambiguities in relating the observable quantities to the physical quantities computed in the two frames~\cite{Faraoni:1998qx,1994-Magnano.Sokolowski-PRD,Brooker:2016oqa,Jarv:2016sow,Kuusk:2016rso}.  

 In many cases, the action in the conformally transformed frame contains the non-canonical kinetic term, and it negates the advantage of transforming to Einstein frame (See Sec. II).  Recently, the current authors have devised an analytical method to study the background equations~\cite{mathew2016low}, and the first order perturbed equations in the Jordan frame~\cite{Mathew:2017lvh}. The approach reduces the higher order perturbed equations to a set of second order differential equations and provides a way to relate the derived quantities to the observed quantities directly. The method was applied to a simple case of $f(\phi) (R + \alpha R^2)$~\cite{Mathew:2017lvh}. 

In this work, we use the above analytical approach to investigate a more non-trivial modified gravity model given by
\begin{equation}
\label{eq:fRphi}
f(R,\phi)=R \exp\left[h(\phi) R \right].
\end{equation}
Here, we have considered a simple but non-trivial model of exponential gravity.
There are two reasons for the above choice: First, in the case of FRW metric 
with $a(t) \propto t^n$, the Ricci scalar goes as:
\begin{equation}
R \sim {n(2n-1)}/{t^2}
\end{equation}
Let us consider the following action:
\begin{equation}
   f(R)=R+\sum_{m \geq 1}\dfrac{\alpha_m}{M_{\rm Pl}^{2m}} R^{m+1}
\end{equation}   
where $\alpha_m$ are dimensionless constants. The ratio between the  $(m+1)^{\rm th}$ term and $m^{\rm th}$ term in the above action is the same for all values of $m$, i. e. ${n(2n-1)}/{(M_{\rm Pl}^2 \, t^2)}$.
Thus, it is not sufficient to include only finite number of higher order Ricci scalar terms.  Second, such a scenario comes from the fact that the quantum corrections to the gravity and scalar field can have a scale dependent corrections~\cite{Donoghue:1993eb,Donoghue:1997hx,Hamber:1995cq,Barth:1983hb} and the series expansion of the above exponent leads to 
\begin{equation}
R \exp\left[h(\phi) R \right] = R + h(\phi) R^2 +\dfrac{1}{2}h(\phi)^2 R^3+...
\end{equation}
Thus, the modification to general relativity appears as a non-minimal coupling 
term to the higher order terms only and \emph{not} to the Einstein-Hilbert action. Since the Einstein-Hilbert action does not contain any non-minimal coupling term, there is no motivation for the conformal transformation. Thus, we can not rely on the conformal transformation techniques used in the literature, and we use the new analytical technique developed in Ref.~\cite{Mathew:2017lvh}. 

Recently there has been great interest in exponential gravity~\cite{Linder:2009jz,Odintsov:2017qif,Xu:2012up,Cognola:2007zu} (see References.~\cite{capozziello2010beyond,Nojiri:2010wj,Nojiri:2017ncd} for comprehensive reviews on modified gravity).  There has been successful attempts to unify inflation and late time acceleration within the context of exponential gravity~\cite{Elizalde:2010ts}. Unlike in the earlier references, we have explicitly obtained an analytical solution during the entire inflationary epoch and have shown there is an exit from inflation in the exponential gravity model.  We have obtained the dependence of the Number of e-foldings with potential. To our knowledge, there is no explicit analytical expression for such inflationary model. It is important to note that the above model contains all the strong gravity corrections and is a non-perturbative result. As mentioned above, since the non-minimal coupling is with all higher-order Ricciscalar terms and not with the Einstein-Hilbert action terms, we have evaluated the perturbations in the Jordan frame. Such a calculation has not been done earlier. Another important feature of our model is that we consider the inflaton to be a scalar field with quartic potential, i.e., the scalar field that drives inflation in our model is compatible with the standard model of particle physics.

We would like to note that inflation is studied with non-minimal inflaton-gravity coupling in Ref.~\cite{Odintsov:2018ggm}. In the paper, the authors have presented an interesting reconstruction method to obtain slow-roll inflationary models. They have considered non-minimal coupling of the inflaton with the Einstein-Hilbert action. However, in our model we obtain exact solutions of inflationary evolution with an exit and also we have considered non-minimal coupling to the higher order terms.

The paper is organized as the following. In the next section, we introduce the model in detail and show that it is convenient to study the model in the Jordan frame. We then obtain the exact analytical inflationary solution exit. In section (\ref{sec:firstorder}), we compute the scalar and tensor power spectra. In the last section, we present the results and conclusions.

We use (-,+,+,+) metric signature and natural units, $c=1,\quad \hslash=1$ and $\kappa\equiv\frac{1}{M_{Pl}^2}$, where $M_{Pl}$ is the reduced Planck mass. Lower Latin alphabets denote the 4-dimensional space-time, and lower Greek letters are used for the 3-dimensional space. Dot represents the derivative with respect to cosmic time $t$, while prime denotes derivative with respect to conformal time $\eta$.  $H\equiv\frac{\dot{a}}{a}$ is the Hubble parameter.

\section{ MODEL AND BACKGROUND SOLUTION}
\label{sec:Model}

We consider the following action in Jordan frame:
 \begin{equation}\label{eq:action}
 S_{J}=\int d^{4}x\sqrt{-g}\left[\frac{1}{2 \kappa } f(R,\phi)- \frac{\omega}{2}   g^{a b} \nabla_{a} \phi \nabla_{b} \phi -V(\phi)\right]  ,
 \end{equation}
where $V(\phi)$ is the scalar field potential and $f(R, \phi)$ is taken to be
 \begin{equation}
\label{eq:Re^hR}
f(R,\phi)=Re^{h(\phi)R},
\end{equation}
and $h(\phi)$ is the scalar field coupling function. Note that 
$\omega$ in Eq. \ref{eq:action} can take either +1(for
canonical scalar field) or -1 (for Ghost scalar field). For the calculations in this work, we set $\omega = +1$, denoting a canonical scalar field.

 The physical motivation for such a scenario comes from the fact that the quantum corrections to the gravity and scalar field can have scale dependent corrections~\cite{Donoghue:1993eb,Donoghue:1997hx,Hamber:1995cq,Barth:1983hb}.
\subsection{Conformal transformation in $f(R,\phi)$ models}
\label{secA}
As mentioned in the Introduction, in the case of Einstein gravity with non-minimally coupled scalar field, or $f(R)$ models, field equations and equations of motions are of higher order~\cite{mukhanov1992theory,tsujikawa,kaiser2010conformal}. One way to get around this problem is to do a conformal transformation to bring the action to the Einstein frame. For Einstein gravity with the non-minimally coupled scalar field, this along with the redefinition of the scalar fields leads to Einstein gravity with minimally coupled scalar fields \cite{kaiser2010conformal,Kallosh:2013maa}. However in our model, or generally in $f(R,\phi)$ models, this is not the case \cite{Abediabbassi}.  

To see this, the conformal transformation  $\tilde{g}_{\mu \nu}=F g_{\mu \nu}$ 
on the above action, leads to: 
 \begin{eqnarray}
 \tilde{S}_{E} &=& \int d^{4}x\sqrt{-\tilde{g}}\left[\dfrac{\tilde{R}}{2\kappa}-\dfrac{1}{2e^{\sqrt{\frac{2\kappa}{3}}\zeta}}\tilde{g}^{a b}\partial_{a}\phi\partial_{b}\phi \right. \nonumber \\
&-& \left. \dfrac{1}{2}\tilde{g}^{a b}\partial_{a}\zeta\partial_{b}\zeta-\tilde{W} \right],
\label{eq:Einaction}  
 \end{eqnarray}
where 
\[
 F = \frac{\partial f}{\partial R} \, , \quad \zeta=\sqrt{\dfrac{3}{2 \kappa}} \ln F\, , \quad \tilde{W}=\frac{FR-f}{\kappa F^{2}}+\frac{V}{F^{2}} \, .
\]

As it is evident, the conformal transformation leads to \textit{non-canonical kinetic term}, and the potential takes a non-trivial form. It should also be noted that the method \cite{mathew2016low} that we use to obtain the exact solution is much more straightforward in Jordan frame. For these reasons, we do all the calculations in the Jordan frame. 

Field equations and equation of motion of the scalar field for a general $f(R,\phi)$ model in the Jordan frame are given by \citep{hwang2000conserved}
\begin{eqnarray}
\label{eq:phijordan}
& & \Box\phi+\dfrac{1}{2}\left(f_{,\phi}-2V_{,\phi}\right)=0  \\ 
\label{eq:fieldeqjordan}
& & FG^{p}_{q}=\left(\phi^{;p}\phi_{;q}-\dfrac{1}{2}\delta^{p}_{q}\phi^{;c}\phi_{;c}\right)\nonumber \\ 
& & \qquad - \dfrac{1}{2}\delta^{p}_{q}\left(RF-f+2V \right)+F^{;p}_{;q}-\delta^{p}_{q}\Box F
\end{eqnarray}

In the rest of this section, we obtain an exact, analytical inflationary solution with exit.
 
\subsection{ Background inflationary solution}
\label{Sec:BGS}
Consider the spatially flat FRW metric given by the line element
\begin{equation}
ds^{2}=-dt^{2}+a(t)^{2}\left( dx^{2}+dy^{2}+dz^{2}\right) ,
\end{equation}
where $a(t)$ is the scale factor. For this background, the field equations (\ref{eq:fieldeqjordan}) and equation of motion of $\phi$ (\ref{eq:phijordan}) 
for the action (\ref{eq:action}) become:
\begin{eqnarray}
\label{eq:feq00}
& &  \dot{\phi}^{2} + 6 F \left(\dot{H} + H^2 \right) 
- 6 \dot{F} H  + (2 V - f) = 0, \\
\label{eq:feqii}
\!\!\!\! & & \!\!\!\! 4 F H^2 + 2 F \left(\dot{H} + H^2 \right) -  \dot{\phi}^2-2 \ddot{F} -4 \dot{F} H  + 2 V  - f = 0~~~~ \\
\label{eq:eomphi}
& &  \ddot{\phi} + 3 H \dot{\phi}  -\left(\frac{f_{\phi}}{2}- V_{\phi}\right) = 0,
\end{eqnarray}
 
 where $F=\frac{\partial f}{\partial R}$.

For the $f(R,\phi)$ model given by Eq.(\ref{eq:Re^hR}), Field equations and equation of motion of $\phi$ become
\begin{eqnarray}
-\frac{1}{2}\dot{\phi}^2 + \frac{3}{\kappa} e^{hR}H^2 - \frac{3}{\kappa}hR e^{hR}H^2 - \frac{3}{\kappa}Rhe^{hR}\dot{H}  +\frac{6}{\kappa}H e^{hR} \left({hR}\right)^{.} &&  \nonumber \\
-V + \frac{3}{\kappa}HhR \left({hR}\right)^{.} &=& 0 \label{eq:feq002}\\
\frac{3}{\kappa}H^2e^{hR}-\frac{3}{\kappa}H^2hRe^{hR}+\frac{2}{\kappa}\dot{H}e^{hR}-\frac{1}{\kappa}\dot{H}hRe^{hR}-V+\frac{1}{2}\dot{\phi}^2 +\frac{1}{\kappa} (e^{hR}(1+hR)\ddot{)} && \nonumber\\
+ \frac{2}{\kappa}H(e^{hR}(1+hR)\dot{)} &=& 0\\
- \dot{V} + \frac{6}{\kappa}H^2 e^{hR}\left({hR}\right)^{.}+\frac{3}{\kappa} e^{hR}\dot{H} \left({hR}\right)^{.} - \frac{12}{\kappa} H \dot{H}hR e^{h(\phi)R} & & \nonumber\\
- \frac{3}{\kappa}\ddot{H}hR e^{hR} -\dot{\phi}\ddot{\phi} - 3H\dot{\phi}^2 &=& 0 \label{eq:eomphi2}
\end{eqnarray} 
where  $h=h(\phi)$ is the coupling function. We would like to make the following points. First, when the scalar field dominates as in the inflationary phase, the evolution of $a(t)$, $h(t)$ and $\phi(t)$ is determined by solving the above set of equations. 
In the radiation dominated era, for which $R = 0$, the scalar field equation (\ref{eq:eomphi2}) satisfies:
\begin{equation}
\ddot{\phi} + 3 H \dot{\phi} + V_{,\phi} = 0
\label{eq:EOMGR}
\end{equation}

and it is identical to the standard General relativity.

\subsubsection{Exact analytical solution}

As in GR, only two of the above differential equations are independent. Unlike 
GR, we have four independent functions --- $a(t), \phi(t), h(\phi), V(\phi)$. To obtain an exact solution, we need to assume the functional form of two of the above functions. We have used physical conditions to fix the two variables. We have 
obtained the exact inflationary solution using two different approaches and both give identical results. First is to assume the form of $a(t)$ and $V(\phi) = m^2 \phi^2 + \lambda \phi^4$. Second, which we discuss below, is to assume of form of $a(t)$ and $h(\phi(t))$. 

In order to keep the calculations tractable and transparent, we make the following ansatz for the scale factor: 
\begin{equation}
\label{eq:Hphi}
  a=a_0e^{-(t_0/t)^2}
\end{equation}
where $a_0$ and $t_{0}$ are constants. $t_0$ will be determined by the parameters of the model. We also choose $h(\phi)R$ to be independent of time. Note that this choice does not assume that $h(\phi)$ and $R$ are independently constants. As we show below, this will lead to non-trivial functional form of $h(\phi)$.

Substituting the above ansatz in Eqs. (\ref{eq:feq00}, \ref{eq:feqii}, \ref{eq:eomphi}) and solving the equations, we obtain the following functional form for the coupling function: 
\begin{equation}
\label{eq:h}
h(\phi) = \dfrac{288 \, e^3 \, t_0^2 \,M_{Pl}^6}{\phi^6-18 \, e \, M_{Pl}^2 \, \phi^4} \, ,
\end{equation}
$\phi(t)$ is given by
\begin{equation}
\label{eq:phi}
\phi=\dfrac{\phi_0}{t},\quad \mbox{where} \quad 
\phi_0=\sqrt{24e} \, t_0 M_{Pl}.
\end{equation}
and the scalar field potential is
\begin{equation}
\label{eq:potential}
V(\phi) = \lambda\phi^4\,  ,
\end{equation}
where $t_0$ is determined by $\lambda$ using the following relation:
\begin{equation}
t_0^2 =\dfrac{1}{96\, e \, \lambda \, M_{Pl}^2}.
\label{eq:t0lambda}
\end{equation}
Note that the parameter $t_0$ is determined by the coupling constant $\lambda$. 

Following are the important features of the about exact analytical 
solution: 
First, this model has an inflationary solution with exit of inflation. It describes the accelerated expansion of the universe for $t < t_f$ where 
$t_f=\sqrt{2/3}t_0$, and the exit happens at $t=t_f$. 
Second,  the number of e-foldings between a given time $t$ and the exit of inflation  depends on the value of $\lambda$ and hence on the value of $t_0$.
\begin{equation}
 \label{eq:Nstar}
 N=\left(\dfrac{t_0}{t}\right)^2-\left(\dfrac{t_0}{t_f}\right)^2=\left(\dfrac{t_0}{t}\right)^2-\dfrac{3}{2}.
\end{equation}
And the scalar field $\phi$ can be expressed in terms $N$
\begin{equation}
\label{eq:phiVstar}
 \phi=\sqrt{24e}\left(N+\frac{3}{2} \right)^{\frac{1}{2}} M_{Pl},
\end{equation}
Third,
the model supports inflationary solution with exit for a range of values of $t_0$. Fourth, since the scalar field decays with time $t$, 
the modified gravity model only provides changes in the high-energy. In the low-energy limit, the model is identical to General relativity. 
Fifth, from Eq.(\ref{eq:h}), it appears that the coupling function has a singularity after the exit of inflation at $\phi=\sqrt{18e}M_{Pl}$. However, since there is no discontinuity in the evolution and Ricciscalar approaches zero (radiation dominated) much before this point, the limits are well defined. Hence, though the coupling function is singular the field equations are well defined.  Also, during radiation dominated era which follows after the exit from inflation 
the scalar field satisfies Eq.(\ref{eq:EOMGR}). 

 Finally, it is known that conformal transformation may not lead to the system with same qualitative properties \cite{Capozziello:2006dj,Bahamonde:2017kbs}.  It is interesting to note that our model provides an equivalent evolution of the Universe in the Einstein frame however with a complicated scalar field potential. The scale factor and time in Jordan frame and Einstein frame are related by the expressions
\begin{equation}
a_E=\sqrt{F}a, 
\end{equation} 
and 
\begin{equation}
t_E=\int_{ti}^{t}\sqrt{F}dt
\end{equation}
For our model, $F=e^{hR}+hR e^{hR}=2e$ is a constant. This means that the time and scale factor in Einstein frame are same as in Jordan frame but scaled by a constant. However, the scalar field potential is highly non-trivial after the conformal transformation as discussed in Sec.~(\ref{secA}).
\begin{figure}[!h]
 \centering
 \begin{minipage}{0.5\textwidth}
        \centering
        \includegraphics[width=0.9\textwidth]{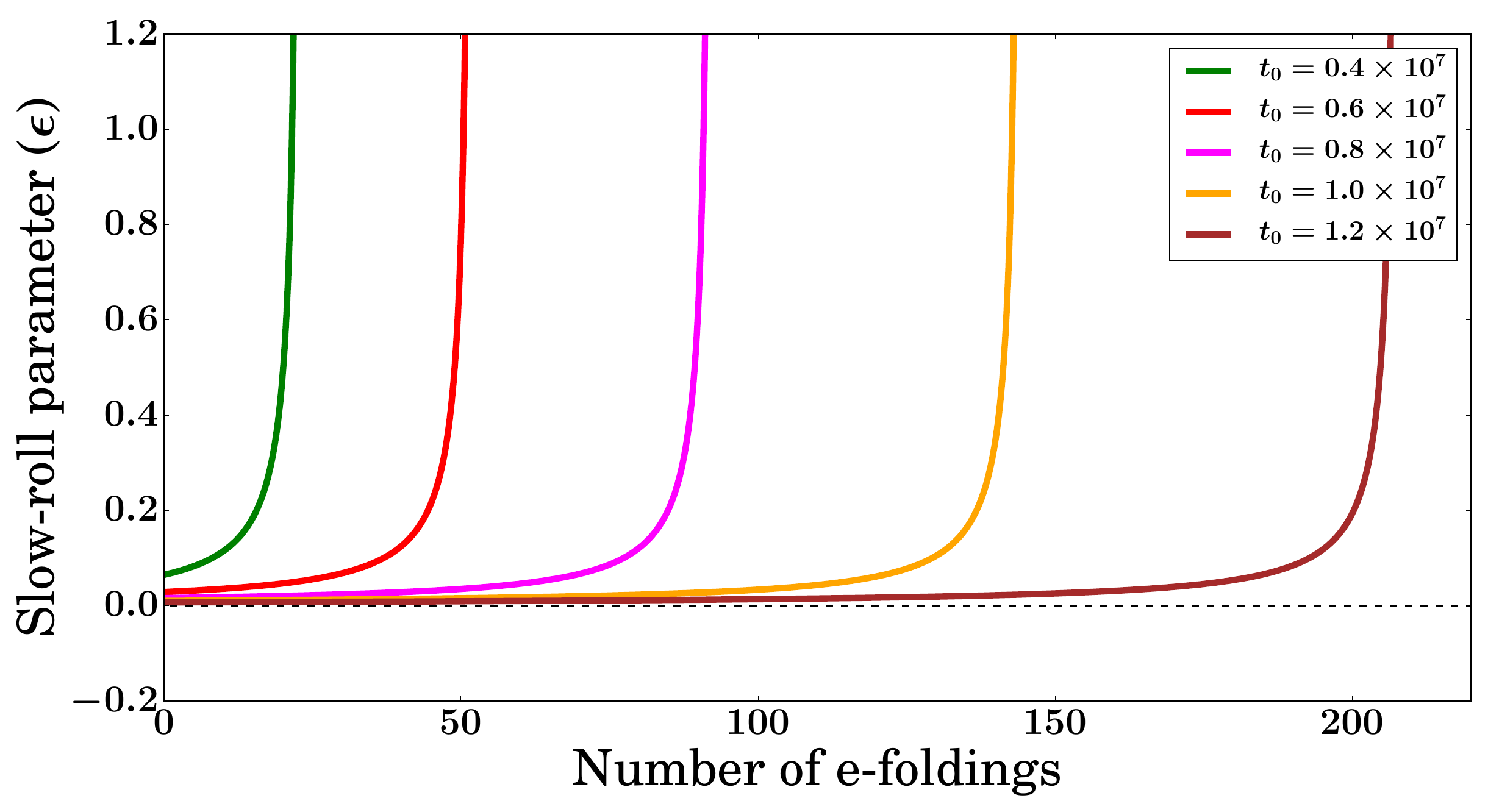} 
       
    \end{minipage}\hfill
    \begin{minipage}{0.5\textwidth}
        \centering
        \includegraphics[width=0.9\textwidth]{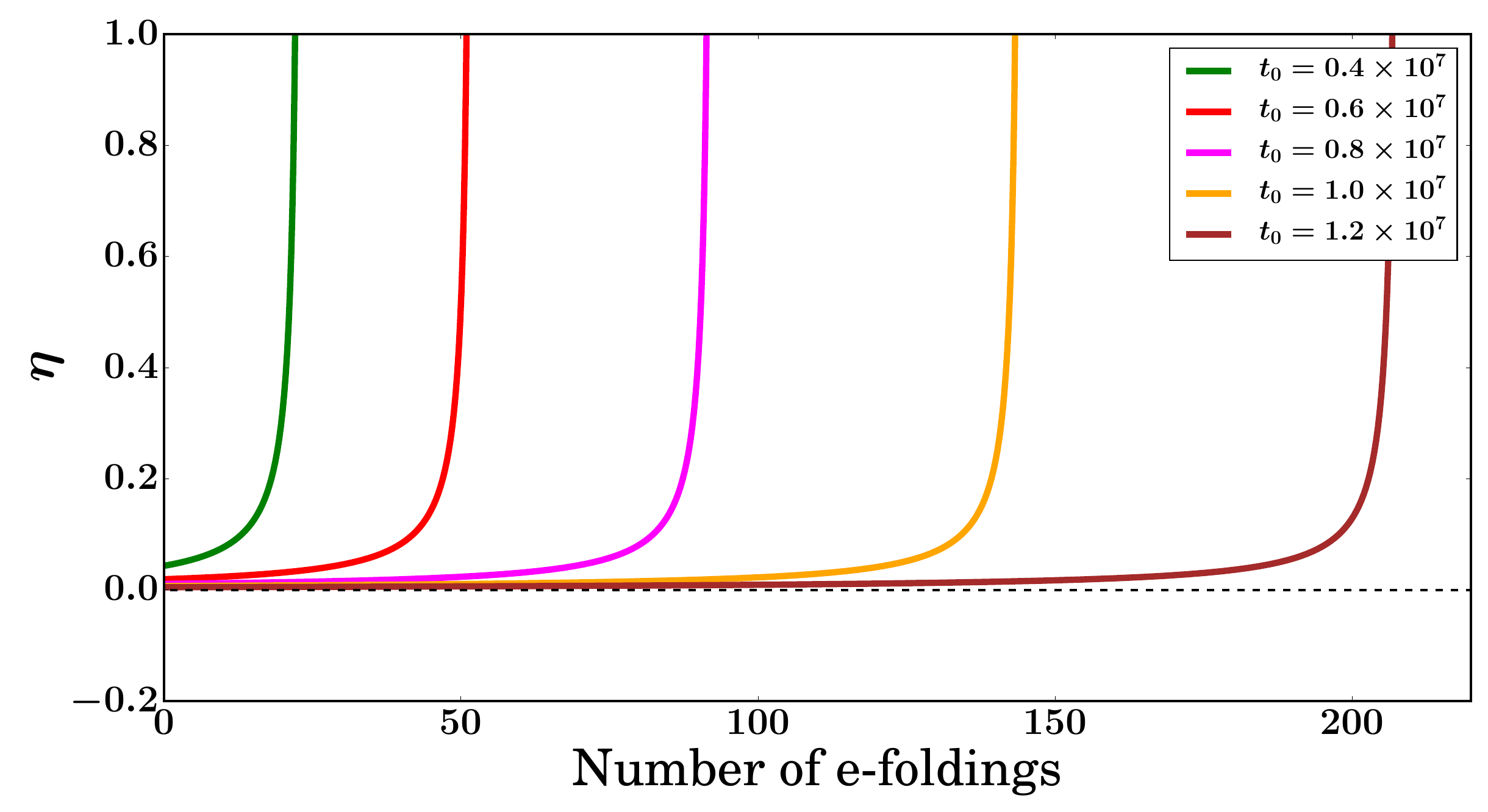} 
    \end{minipage}
  \caption{Plot of the slow-roll parameters $\epsilon(= -\frac{ \dot{H}}{H^2})$ and $\eta(=-\frac{\dot{\epsilon}}{\epsilon H})$, for different values of $t_0$, as a function of the number of e-foldings.}
 \label{fig:epsilon-etavsN}
 \end{figure}
\begin{figure}[!h]
 \centering
 \begin{minipage}{0.5\textwidth}
        \centering
        \includegraphics[width=1\textwidth]{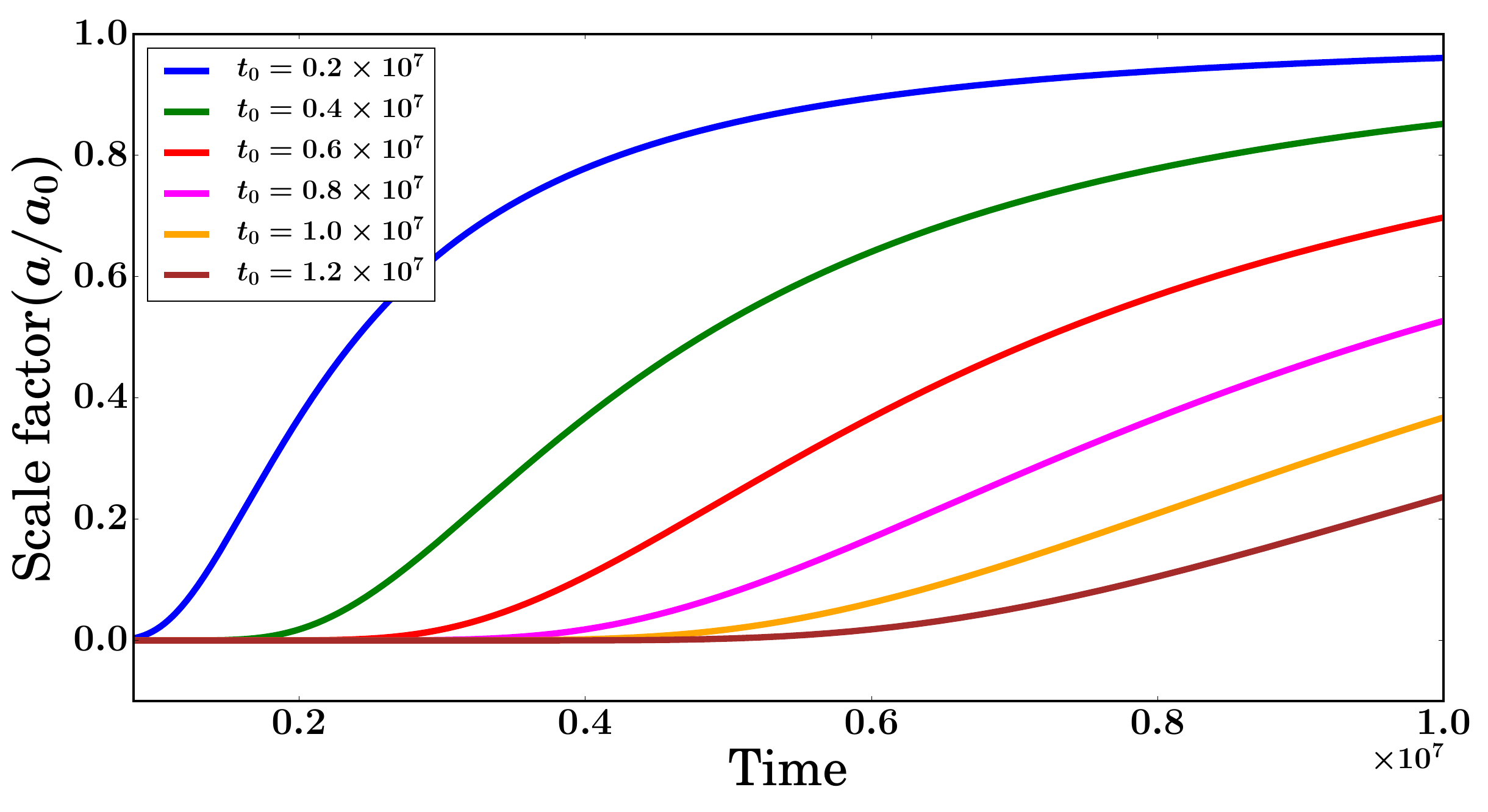} 
       
    \end{minipage}\hfill
    \begin{minipage}{0.5\textwidth}
        \centering
        \includegraphics[width=1\textwidth]{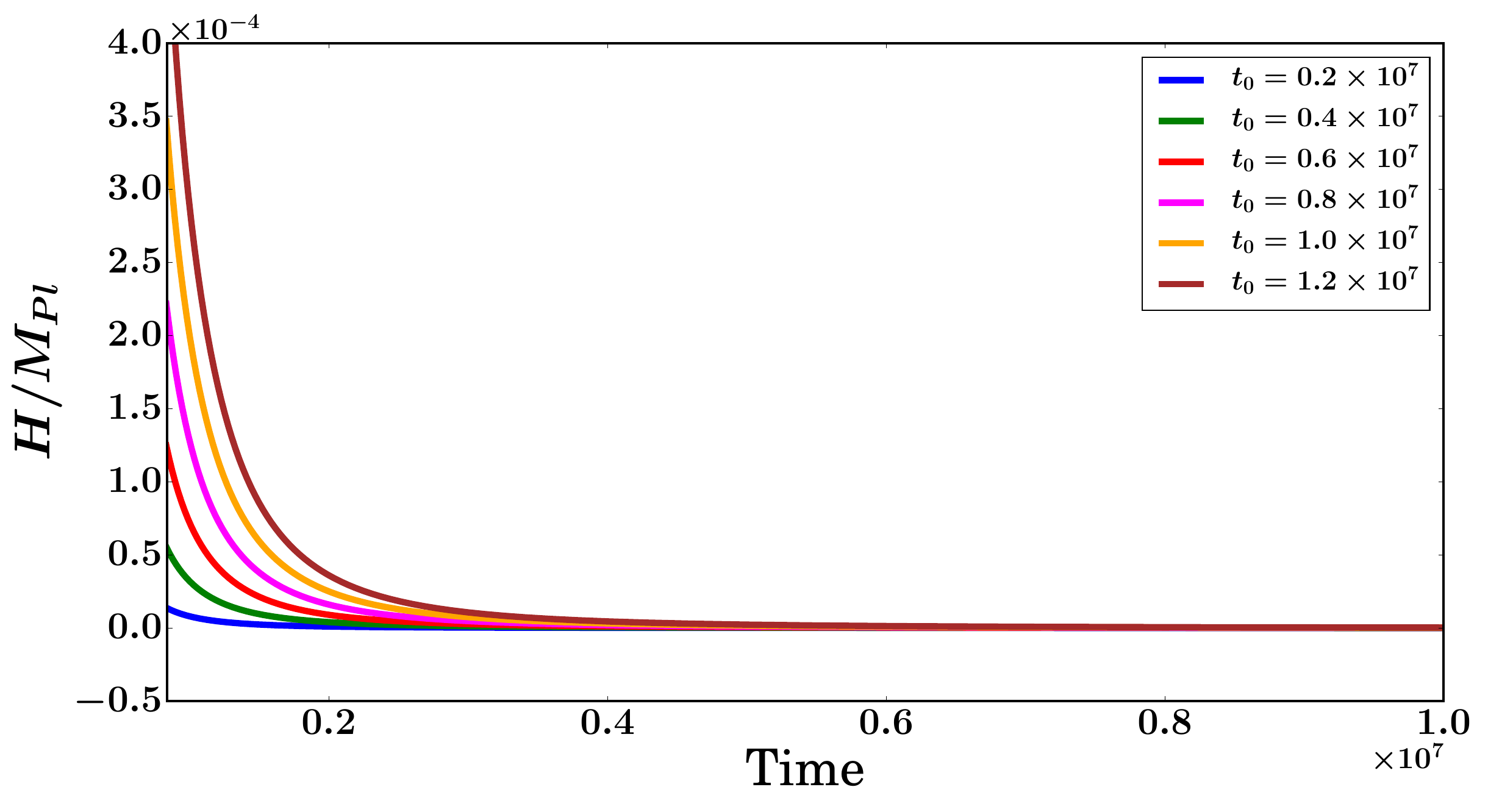} 
    \end{minipage}
  \caption{Scale factor and Hubble parameter vs. time. (i) Scale factor $(a/a_0)$ and (ii) Hubble parameter $(H/M_{Pl})$ for different values of $t_0$.}
 \label{fig:Hubblevst}
\end{figure}


In Figure 1, we have plotted the slow-roll parameters as a function of Number of e-foldings for different values of $t_0$. 
In Figure 2. we have plotted the scale factor and the scaled Hubble parameter as a function of cosmic time $t$.  The plots clearly show the inflationary phase and the exit from inflation. We would like to point that we have plotted for different values of $t_0$ around
$t_0 = 8.41 \times 10^6 M_{Pl}^{-1}$. These values of $t_0$ are obtained by using the constraint on inflationary energy scale fixed by the Planck data ~\cite{Akrami:2018odb}
\begin{equation}
\label{eq:Vconstraint}
V<(1.7 \times 10^{16} GeV)^4 \, .
\end{equation}

\section{First order scalar and tensor perturbations}
\label{sec:firstorder}
The first order perturbations about the FRW background is given by
\begin{eqnarray}
\label{eq:pertubed_le}
ds^2 &=&-(1 + 2\theta)dt^2-a(\beta_{,\alpha} + B_\alpha )dtdx^{\alpha} \nonumber \\
&+& a^2 [g^{(3)}_{\alpha\beta} (1-2\psi) + 2\gamma _{,\alpha|\beta} + 2C_{\alpha|\beta} + 2C_{\alpha\beta} ].
\end{eqnarray}
where $a(t)$ is the cosmic scale factor with $dt \equiv ad\eta$. $\theta(\textbf{x},t), \beta(\textbf{x},t),\psi(\textbf{x},t),$ and $\gamma(\textbf{x},t)$ denote the scalar perturbation. $B_{\alpha}(\textbf{x},t)$ and $C_{\alpha}(\textbf{x},t)$ are trace-free vector perturbation and $C_{\alpha \beta}(\textbf{x},t)$ is the transverse and trace-free tensor perturbation. The vertical bar denotes the covariant derivative on 3-D spatial hypersurface characterized by $g^{(3)}_{\alpha\beta}$. We decompose the scalar field as $\phi(\textbf{x},t)=\bar{\phi}(t)+\delta\phi(\textbf{x},t)$. 

In the Newtonian gauge, scalar perturbations in the Fourier space satisfies the following equations \cite{hwang2000conserved}:

\begin{eqnarray}
\label{eq:deltaF}
\delta F -F_{\phi}\delta\phi-F_{R}\delta R &=& 0.
\\
\label{eq:inotjNg}
-F\psi +F \theta +\delta F &=& 0.
\\
\label{eq:0iNg}
-2F\dot{\psi} -2 F H \theta -\dot{F} \theta +\dot{\phi} \delta\phi +\dot{\delta F} -H \delta F &=& 0.
\\
\label{eq:00Ng}
6 F H \dot{\psi} + 6 F H^{2}\theta +2 F \dfrac{k^{2}\psi}{a^{2}}-\dot{\phi}^{2}\theta + 3\dot{F}\dot{\psi} + 6 \dot{F}H \theta \nonumber \\ +\dot{\phi} \dot{\delta \phi} - \ddot{\phi} \delta \phi - 3H\dot{\phi} \delta \phi -3 H \dot{\delta F} +3 \dot{H} \delta F + 3 H^{2} \delta F - \dfrac{k^{2}\delta F}{a^{2}} &=& 0.
\\
\label{eq:iiNg} 
6 F \ddot{\psi} +12 F \dot{H} \theta + 6 F H \dot{\theta} +12 H F \dot{\psi} +12 F H^{2} \theta - 2F \dfrac{k^{2}\theta}{a^{2}} + 3 \dot{F} \dot{\psi} + 6 \dot{F}H \theta +3 \dot{F}\dot{\theta} \nonumber\\ +4 \dot{\phi}^{2}\theta + 6 \theta \ddot{F} -4 \dot{\phi}\dot{\delta \phi} -2 \ddot{\phi}\delta \phi - 6H \dot{\phi} \delta \phi -3 \ddot{\delta F} -3 H \dot{\delta F} + 6 H^{2} \delta F -\dfrac{k^{2}\delta F}{a^{2}} &=& 0.
\\
\label{eq:deltaphi}
\ddot{\delta\phi}+3 H\dot{\delta\phi}-\dfrac{1}{2}f_{\phi\phi}\delta\phi + V_{\phi \phi}\delta\phi +\dfrac{k^2}{a^2}\delta\phi -3\dot{\phi}\dot{\psi}-6H\dot{\phi}\theta-\dot{\phi}\dot{\theta
} \nonumber\\ -2\ddot{\phi}\theta+3F_{\phi}\ddot{\psi}+6F_{\phi}\dot{H}\theta
+3F_{\phi}H\dot{\theta}+12F_{\phi}H\dot{\psi}+12F_{\phi}H^{2}\theta
+2F_{\phi}\dfrac{k^2}{a^2}\psi-F_{\phi}\dfrac{k^2}{a^2}\theta &=& 0.
\end{eqnarray}

where
\[
 F_{\phi} = \frac{\partial F}{\partial \phi} \quad \mbox{and} \quad F_{R} = \frac{\partial F}{\partial R} \, .
\]
And the tensor perturbations satisfy the following differential equation\cite{hwang2000conserved}:
\begin{equation}
\label{eq:tenoreq}
\ddot{C}^\alpha_\beta+ \left(\dfrac{\dot{F}}{F}+3H\right)\dot{C}^\alpha_\beta+\dfrac{k^2}{a^2}C^\alpha_\beta=0.
\end{equation}
\subsection{Scalar power spectrum}

In this section, we derive the equation satisfied by the 3-curvature perturbation $\mathcal{R}$ and calculate corresponding power spectrum. We follow the procedure used in Ref.~\cite{Mathew:2017lvh}. In Jordan frame, $\mathcal{R}$ is defined as 
\begin{equation}
\label{eq:cur_per_def}
\mathcal{R}=\psi+\dfrac{H}{\dot{\phi}}\delta\phi.
\end{equation}
%

Since equations governing the scalar perturbations are highly complicated, instead of trying to solve for $\mathcal{R}$ directly, we solve the equations 
for other perturbed quantities and use those solutions to derive the equation satisfied by $\mathcal{R}$.

First, we define new variables so that it will reduce the total no. of variables and make the equations simpler. We define a new variable $\Theta=\theta+\psi$. Using Eqs. (\ref{eq:inotjNg}, \ref{eq:0iNg}, \ref{eq:00Ng}), we get
\begin{eqnarray}
& & F \ddot{\Theta} + \left(-\dfrac{6F\ddot{\phi}}{\dot{\phi}}+4\ddot{F}+2\dot{F}H-4F\dot{H} \right)\theta \nonumber\\ 
\label{eq:Thetatheta}
&+& \left(FH-\dfrac{2F\ddot{\phi}}{\dot{\phi}}+3\dot{F}\right)\dot{\Theta} \\
&+& \left(\dfrac{2\dot{F}\ddot{\phi}}{\dot{\phi}}+4F\dot{H}-\dfrac{2HF\ddot{\phi}}{\dot{\phi}}-\ddot{F}+\dot{F}H+F\dfrac{k^2}{a^2}\right)\Theta = 0.
\nonumber
\end{eqnarray}

Substituting the background quantities corresponding to the exact solution obtained in the previous section, in the sub-Hubble scales, $\Theta$ satisfies the following differential equation:
\begin{equation}
\label{eq:Thetaeq}
\ddot{\Theta}+\left(\dfrac{2 \, t_0^2}{t^3}+\dfrac{4}{t}\right)\dot{\Theta}+\dfrac{k^2}{a^2}\Theta=0.
\end{equation}

Rewriting $\delta\phi$ in terms of $\Theta$, we get,
\begin{equation}
\label{eq:deltaphiexp}
\delta \phi=-\dfrac{1}{\sqrt{6e} \, t_0}\dfrac{(\dot{\Theta} \, t^3+2 \, t_{0}^2 \, \Theta)}{t}.
\end{equation}
{
In the perturbed equations one can replace all the perturbed quantities with $\Theta$, $\mathcal{R}$ and their derivatives. Also using Eq.~(\ref{eq:Thetaeq}) and its derivatives, one can write the higher derivatives of $\Theta$ in terms of $\Theta$ and $\dot{\Theta}$. Now eliminating $\Theta$ and $\dot{\Theta}$ from the perturbed equations we obtain the equation.
\begin{eqnarray}
\label{eq:cpeqold}
& &\ddot{\mathcal{R}}\left(1+\mathcal{O}\left(\frac{1}{k^2}\right)\right) +\left(\dfrac{6 t_0^2}{t^3}+\dfrac{2}{t}\right)\dot{\mathcal{R}}\left(1+\mathcal{O}\left(\frac{1}{k^2}\right)\right)+\nonumber \\& & \dfrac{k^2}{a^2}\mathcal{R}\left(1+\mathcal{O}\left(\frac{1}{k^2}\right)\right)=0.
\end{eqnarray}}
 
{The above equation, valid in sub-horizon scale, can be simplified to}
\begin{equation}
\label{eq:cpeq}
\ddot{\mathcal{R}}+\left(\dfrac{6 \, t_0^2}{t^3}+\dfrac{2}{t}\right)\dot{\mathcal{R}}+\dfrac{k^2}{a^2}\mathcal{R}=0.
\end{equation}
This is the second key result of this work regarding which we would like to mention the following: 
First, the procedure allows us to separate the higher order differential equation into two second order equations. 
In this case, we have evaluated the evolution of the adiabatic curvature perturbation ${\cal R}$. Second, the other second order differential equation 
corresponds to the evolution of the isocurvature perturbations~\cite{Gordon:2000hv}. 
We have ignored the contribution of isocurvature perturbations.

In the short wavelength limit, Eq.~(\ref{eq:cpeq}) can be rewritten as
 \begin{equation}
\label{eq:nus}
\nu''+k^2 \nu=0,
\end{equation}
Where \textit{prime} denotes derivative with respect to conformal time, $\mathcal{R}= \nu/z$, and $z=a \, t$. Using the Bunch-Davies vacuum at the initial epoch of inflation, in the sub-Hubble scales, the curvature perturbation is given by
\begin{equation}
\label{eq:Rshort}
\mathcal{R}_{<}=\dfrac{1}{z\sqrt{2k}}e^{-ik\eta}.
\end{equation}
Matching this solution with the super-Hubble scale solution at the Hubble exit ($k=aH$)
\footnote{To simplify the calculations, we have used power-law fit for the term $aH$}
, the scalar power spectrum is given by
\begin{equation}
\label{eq:scalarps}
\mathcal{P}_{\mathcal{R}}=k^3|\mathcal{R}|^2= \mathcal{A}_s k^{n_s-1},
\end{equation}
where $n_s$ is the scalar spectral index and the value of $\mathcal{A}_s$ depends on $\lambda$. 

The value of scalar spectral index $n_s$ depends only on the number of e-foldings ($N$) at which the power spectrum is evaluated. Table (\ref{tab:Nns})below list various values of $N$ and the corresponding values of scalar spectral index $n_s$.

\begin{center}
\begin{table}[!h]
\renewcommand\arraystretch{1} 

\begin{tabular}{ |m{3cm}|m{3cm}| } 

 \hline
\qquad$N$ & \qquad$n_s$\\  
 \hline
 \qquad 40  & \qquad 0.907  \\ 
 \hline
 \qquad 60   & \qquad 0.939  \\ 
 \hline
 \qquad 80   & \qquad 0.956  \\ 
 \hline
 \qquad 90   & \qquad 0.961  \\ 
 \hline
 \qquad 100   & \qquad 0.965  \\ 
 \hline
 \qquad 110   & \qquad 0.968  \\ 
 \hline
 \qquad 115   & \qquad 0.969  \\ 
 \hline

\end{tabular}
 \caption{Value of scalar spectral index $n_s$ for various values of number of e-foldings $N$}
 \label{tab:Nns}
 \end{table}
 
\end{center}
Here we see that for $90 \leq N \leq 115$, value of scalar spectral index matches with the observational data $n_s=0.965 \pm 0.004$ \cite{Akrami:2018odb}. In the next subsection, we calculate the tensor power spectrum and obtain the value of tensor spectral index for the above range of $N$.

\subsection{Tensor power spectrum}
Substituting the background quantities, Eq. (\ref{eq:tenoreq}) takes the following form: 
\begin{equation}
\label{eq:tensorpert}
\ddot{C}^\alpha_\beta+ \dfrac{6 \, t_0^2}{t^3}\dot{C}^\alpha_\beta+\dfrac{k^2}{a^2}C^\alpha_\beta=0.
\end{equation}

We can simplify this equation by rewriting $C^\alpha_\beta=\nu_g/z_g$, where $z_g=a$. Eq.~(\ref{eq:tensorpert}) becomes,
\begin{equation}
\label{eq:nueq}
\nu_g''+\left(k^2 -\dfrac{z_g''}{z_g}\right)\nu_g =0.
\end{equation}

 At short wavelengths, solution to the above equation is given by
 \begin{equation}
 \label{eq:nusolution}
 \nu_g=\dfrac{1}{\sqrt{2k}}e^{-ik\eta}.
 \end{equation}
The tensor power spectrum is given by
 \begin{equation}
 \label{eq:tesorps}
 \mathcal{P}_g=\mathcal{A}_t \, k^{n_t},
 \end{equation}
where $n_t$ is the spectral index and the value of $\mathcal{A}_t$ depends on the value of $\lambda$. The values of the tensor spectral index $n_t$ for  $90 \leq N \leq 115$ are given in Table (\ref{tab:Nnt}).
Here we see that the values of tensor spectral index $n_t$ are negative, indicating a spectrum with red tilt.

\begin{center}
\begin{table}[!h]
\renewcommand\arraystretch{1} 

\begin{tabular}{ |m{3cm}|m{3cm}| } 

 \hline
\qquad$N$ & \qquad$n_t$\\  
 \hline
 \qquad 90  & \qquad -0.029  \\ 
 \hline
 \qquad 95   & \qquad -0.028  \\ 
 \hline
 \qquad 100   & \qquad -0.026  \\ 
 \hline
 \qquad 105  & \qquad -0.025  \\ 
 \hline
 \qquad 110   & \qquad -0.024  \\ 
 \hline
 \qquad 115   & \qquad -0.023  \\ 
 \hline
 
\end{tabular}
 \caption{Value of tensor spectral index $n_t$ for various values of number of e-foldings $N$}
 \label{tab:Nnt}
 \end{table}
 
\end{center}

\section{RESULTS AND CONCLUSION}
In this work, we proposed an inflationary model in which a self-interacting massless scalar field is non-minimally coupled to $f(R)$ gravity. 
Even though it is common to study the $f(R)$ models in the Einstein frame, we showed explicitly that in the Einstein frame, 
action contains the non-canonical kinetic term. In our case, we performed all the analysis including the first order 
perturbation analysis in the Jordan frame. 
As mentioned in the introduction there is great interest in exponential gravity with the possibility of connecting the early Universe inflationary phase with late Universe dark-energy phase~\cite{Elizalde:2010ts,Odintsov:2018qug}. The difference between these early works and ours is that the coupling constant in the exponential
gravity is not a constant; it is dynamical. In reference \cite{Elizalde:2010ts}, the authors need to fix the
value of the cosmological. In reference \cite{Odintsov:2018qug}, the authors require four additional parameters 
$(b, c, \gamma_0, \gamma_1)$ to determine the evolution from the early Universe to Inflation. In our model, 
there is no free parameter; the only unknown parameter is fixed with the start of inflation. The key advantage
is that the coupling constant naturally leads to GR after inflation without any fine-tuning.

We explicitly showed that the model supports inflationary solution with exit for a range of values of $t_0$, which is determined by the parameter
$\lambda$. 
We obtained first order scalar and tensor perturbation equations explicitly. 
We have used the new analytical method devised in Ref. \cite{Mathew:2017lvh} to reduce the higher order scalar perturbation equations to second order 
in 3-curvature perturbation in the short wavelength limit.  We analytically calculated the scalar power-spectrum and 
obtained the range on the number of e-foldings which corresponds to the acceptable value of the scalar spectral index.
We obtained the tensor power-spectrum and showed that the tensor power spectrum is red tilted. 

We plan to numerically evolve scalar and tensor perturbations and evaluate the amplitude of scalar and tensor fluctuations and constrain the
numerical values of the parameters. We also plan to extend the analysis to the higher order, specifically to evaluate the non-Gaussianity 
and find the distinguishing features between inflationary models in modified gravity and general relativity. 

\section{Acknowledgements}
We thank Debottam Nandi for fruitful discussions and suggestions. JM thank Prof. L. Sriramkumar for supporting him during his visit at IITM.
JPJ is supported by CSIR Junior Research Fellowship, India and JM was supported by UGC Senior Research Fellowship, India. The work is supported by DST-Max Planck Partner Group on Gravity and Cosmology 



\bibliography{bibdata.bib}

\end{document}